\documentclass[twocolumn,showpacs,preprintnumbers,amsmath,amssymb]{revtex4} 
\usepackage{graphicx}
\usepackage{dcolumn}
\usepackage{bm}

\def\ni{\noindent}

\def\beq{\begin{equation}} 
\def\eeq{\end{equation}}

\begin{document} 
 
 
\title{Da Vinci Fluids, catch-up dynamics and dense granular flow}

\author{Raphael Blumenfeld}
\email[]{rbb11@cam.ac.uk}

\affiliation{Earth Science and Engineering, Imperial College, London SW7 2AZ, UK}
\altaffiliation{Also at: Cavendish Laboratory, JJ Thomson Avenue, Cambridge CB3 0HE, UK}

\author{Sam F. Edwards}
\altaffiliation{Cavendish Laboratory, JJ Thomson Avenue, Cambridge CB3 0HE, UK}

\author{Moshe Schwartz}
\affiliation{Beverly and Sackler School of Physics and Astronomy, Tel Aviv University, Ramat Aviv 69934, Israel}

\date{\today}

\begin{abstract} 
We introduce and study a da Vinci Fluid, a fluid whose dissipation is dominated by solid friction. We analyse the flow rheology of a discrete model and then coarse-grain it to the continuum. We find that the model gives rise to behaviour that is characteristic of dense granular fluids. In particular, it leads to plug flow. We analyse the nucleation mechanism of plugs and their development. We find that plug boundaries generically expand and we calculate the growth rate of plug regions. In systems whose internal effective dynamic and static friction coefficients are relatively uniform we find that the linear size of plug regions grows as (time)$^{1/3}$. The suitability of the model to granular materials is discussed. 

\end{abstract} 
\pacs{47.57.Gc, 45.70.Mg, 62.20.Qp}
\keywords{Plug flow, granular fluids, solid friction, da Vinci Fluid}

\maketitle

\section{Introduction}

\ni The flow of non-cohesive granular matter has focused much attention due to both a broad technological relevance and deep scientific challenges. The importance of modeling granular flow cannot be overemphasized - particulate transport is important to dry chemicals, pharmaceutical granules, agricultural grains, cereals, pebble beds in nuclear reactor and more. Of particular interest is flow in hoppers and silos, commonly used for storage and discharge. Dense particulate flows are also relevant to many natural phenomena: avalanches of rocks and snow, cratering, quicksand dynamics and dune locomotion. 

\ni Flow of dilute systems is straightforward to model, using pair inelastic collision theories \cite{Ha83}-\cite{Go99}. In many cases, however, the flow is dense and even the concept of collision is not well-defined when grains are in prolonged contact. Many attempts to model dense flow \cite{Tsimring}-\cite{Mi9900} follow a trend which favors increasing model complexity to accommodate addition of physical mechanisms. While this helps in fitting experimental measurements, it makes models less general and difficult to extract generic insight from. 

\ni Here we focus on a minimal model - a fluid governed solely by solid friction. Solid friction plays a particularly important role in the flow of dense non-cohesive particulates - it provides a mechanism to dissipate mechanical energy as grains rub against neighbours, and to store it in intra-granular degrees of freedom. We call this a da Vinci fluid (dVF), after da Vinci's pioneering work on solid friction \cite{dV}, which preceded the more known works of Amontons \cite{Am1699} and Coulomb \cite{Co1779}. In a dVF, volume elements follow Newton's second law, but they interact via the da Vinci - Amontons - Coulomb law of solid friction. We show that this model captures correctly one of the most important aspects of flow of dense granular materials - the formation and growth of plug flows. We present exact solutions under simplified conditions that provide insight into the effect of this mechanism on the macroscopic rheology. 

\ni In the following we first formulate the flow equations for laminar flow in a model discrete system. We then solve the equations for several simple cases. Next, we use the gained insight to obtain continuum equations. Finally, we derive the dynamics of growth of simple plug regions. 

\ni We construct our model for laminar flow of dense granular material by dividing the system into layers, each of many grains, perpendicular to the direction of flow. Individual grains across the common interface between adjacent layers interact via normal and friction contact forces. These give rise to mean normal and tangential friction forces between the layers. Following experimental and numerical observations \cite{Jackson83}\cite{Jop06}, we assume that the threshold for relative motion between layers occurs once the ratio of the friction force to the normal force exceeds an effective static friction coefficient $\mu_s$. Once relative motion has been established, the friction force is velocity-independent\cite{Co1779} and is proportional to the normal force with a dynamic friction coefficient $\mu_d<\mu_s$. 

\ni The discrete system is shown in figure 1. It comprises a set of $N$ parallel layers of width $d$, all moving in the $z$-direction. The layers are confined between two boundary plates and oriented in the direction of gravity. A pressure $P$ is applied to the boundary plates, as shown in figure 1. Effective dynamic and static friction coefficients are assumed both between the layers and between the outermost layers and the boundary plates. The focus on planar layers is mainly for convenience - the following analysis is general and applies to any uni-directional flow, such as parallel streamlines in a pipe. For simplicity, we assume that the layers have a uniform mass density $\rho$, but the model can be extended to non-uniform density, in which case the local effective friction coefficient due to friction between grains in adjacent layers, depends on the local density and hence must be position dependent. In the present article we consider, however, only cases where the initial density is uniform and remains so throughout the dynamics. More general cases, where the friction coefficient depends on the inertial number $I$ \cite{Jop06}\cite{DaCruz05}\cite{Chevoir07}, will be addressed in a later report. 

\ni When moving relative to the boundaries, the layersÕ motion in the $z$-direction is given by  

\begin{equation}
\rho d \dot{v}_n = \rho d g + 
\begin{cases}
p_{n+1,n} - \mu_d P &if $\ n=1\ $; \cr
p_{n-1,n} + p_{n+1,n} &if $\ 1{$<$}n{$<$}N\ $; \cr
p_{n-1,n} - \mu_d P &if $\ n=N$ \ , \cr
\end{cases}
\label{Ai}
\end{equation}
where $n = 1, 2, ..., N$ indexes the layers and $p_{n+1,n}$ is the friction force per unit area that layer $n+1$ applies on layer $n$. 
Note that the above equations are correct whenever $g>\mu_sP/(\rho d)$, regardless of the relative velocity between the boundary and the outer layers.
By Newton's third law, this force is equal and opposite to the force that layer $n$ applies on layer $n+1$, $p_{n+1,n}=-p_{n,n+1}$. 

\ni The velocity of the centre of mass of all the layers is $v_{CM} = \sum_n v_n / N$ and, by summing the equations, it is straightforward to verify that the acceleration of the centre of mass is 

\begin{equation}
\dot{v}_{CM} = g - \frac{2\mu_d P}{\rho L} \ ,
\label{Aii}
\end{equation}
where $L$ is the distance between the stationary boundary plates. 
The friction forces on the right hand side of eqs. (\ref{Ai}) follow the da Vinci - Amontons - Coulomb law and depend non-analytically on the relative velocity between neighboring layers. 

\ni Noting the significance of plug flow, we wish to determine the consistency of such solutions with the threshold nature of the friction forces. We therefore consider first a case where all the initial velocities and accelerations of the layers are equal, given by eq. (\ref{Aii}). 
Calculating the friction force per unit area between layers $n$ and $n+1$, we find 
\begin{equation}
p_{n+1,n} = \mu_d P\left( 1 - \frac{2n}{N}\right) \ ,
\label{Aiv}
\end{equation}
decreasing linearly with distance from the left boundary. From (\ref{Aiv}), all the inter-layer friction forces are below the threshold value $\mu_s P$, i.e. the layers cannot slide relative to one another and they move together. Thus, there exists a uniform plug flow solution where all the layers fall as one rigid body. 

\ni Next, let every layer move initially relative to both its neighbors, with an overall velocity profile that increases from left to right,
 \begin{equation}
v^0_1 < v^0_2 < ... < v^0_{N-1} < v^0_N  \ .
\label{Avi}
\end{equation}
Since layer 2 falls faster it applies on layer 1 a friction force (per unit area) $p_{2,1}=\mu_d P$ in the $z$-direction. This force is cancelled by the friction force on 1 from the left boundary and 1 accelerates at $a_1 = g$. Similarly, the friction forces on every layer $n<N$ cancel out, leading to all these layers accelerating initially at $g$. Thus, for a short time $t$, before any threshold is reached, the velocity profile for $1\leq n < N$ remains unchanged, $v_{n+1} - v_n =$ constant. 

\ni This solution, however, is unstable. Layer $N$ experiences decelerating friction forces from both layer $N-1$ and the right boundary (figure 2) and its acceleration is $g - 2\mu_d P / \rho d < g$. Therefore, layer $N-1$ gains on layer $N$ and, after a time

\begin{equation} 
\tau_1 = \frac{\rho d(V^0_N - V^0_{N-1})}{2\mu P} 
\label{Avii} 
\end{equation} 
their velocities match at $v(\tau_1)=v^0_{N-1}+g\tau_1$. At that moment $p_{N,N-1}$ vanishes and the two layers move as one, forming a nuclear plug. From $\tau_1$ on, eqs. (\ref{Ai}) need to be solved afresh with a new velocity profile  

\begin{equation} 
v^0_n(\tau_1) = 
\begin{cases}
v^0_n + g\tau_1 &if $\ n{$<$}N\ $; \cr 
v^0_{N-1} + g\tau_1 &if $\ n=N$ \ . \cr
\end{cases}
\label{Aviii} 
\end{equation} 
The acceleration of the nuclear plug is $g - \mu_d P / \rho d < g$. Layer $N-2$ then starts gaining on the plug, catching up with it at $\tau_2=\rho d(v^0_{N-1} - v^0_{N-2})/\mu_d P$. At that moment layer $N-2$ joins the plug and the three layers move as one.  This involves instantaneous readjustment of the friction forces between the layers within the plug: $p_{N,N-1}$ becomes $-\mu_d P/3$ and $p_{N-1,N-2}$  drops below the threshold to $\mu_d P/3$. 
The process of velocity matching and readjustment of inter-layer friction forces in the growing plug region (PR) continues from right to left until it reaches layer 1. From then on, the entire system is a plug, moving as one rigid body. Thus, the flow converges to the first case discussed above. 

\smallskip 
\ni The time interval, $\tau_m$, between the moments that layers $N-m$ and $N-m+1$ join the plug is  

\begin{equation} 
\tau_m = m \frac{\rho d(v^0_{N-m}-v^0_{N-m+1})}{2\mu_d P} 
\label{Aviiia} 
\end{equation} 
and the acceleration of the right $m$-layer plug during this interval is $g-2\mu_d P/m\rho d$. The time it takes the system to converge to a global plug is   

\begin{equation} 
T=\sum_{m=1}^N\tau_m = \frac{\rho d}{2\mu_d P} \sum_{m=1}^N \left(  v^0_m - v^0_1 \right) \ , 
\label{Aviiib} 
\end{equation} 
and then it accelerates at $g-2\mu_d P/\rho L$. The evolution of the velocity profile is shown in figure 1.  

\begin{figure}
\includegraphics[width=4cm]{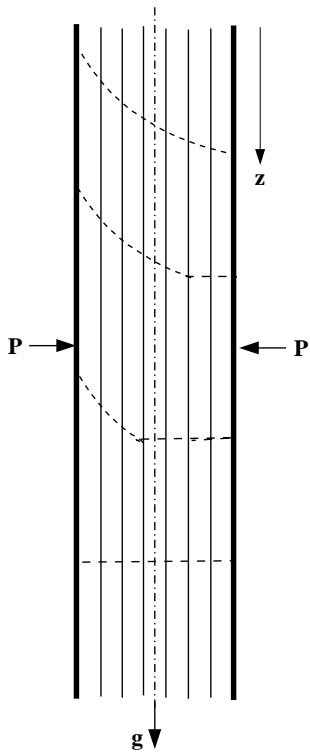}
\caption{\label{Fig2} 
The evolution of a graded velocity profile (dashed line), shown at successive times from top to bottom. A PR forms at the right boundary and expands into the fluid.}
\end{figure}

\ni Next, consider an initial arbitrary non-monotonic velocity profile $v^0_n$, containing no PR. Layers whose velocities are neither a local maximum nor a local minimum of $v^0_n$ move faster than one neighbor and slower than the other. Hence, the friction forces on them cancel out and they accelerate at $g$. 
When the velocity of the layer is a local maximum of the velocity profile (figure 2), both its neighbors slow it down and it accelerates at $g-2\mu_d P$. Similarly, when a layer's velocity is a local minimum, its neighbors act to accelerate it at $g+2\mu_d P$. Both these cases lead to plug nucleation.  
Thus, the catch-up dynamics described above generate nuclei of plugs at all the extrema of $v^0_n$ and these plugs subsequently expand. 

\ni When two expanding PR's meet they merge and continue as one plug. The merging dynamics is interesting in its own right, but it is downstream from this discussion (but see comment below). The expansion and merging of PR's continues until the entire system is one plug. 
Thus, the global plug flow discussed above is a stable fixed-point solution of a whole family of laminar flows. 

\ni The coalescence of PR's is far from simple. When the boundaries of two PR's collide they are at different velocities and accelerations and the velocity matching between them is accompanied by readjustments of inter-layer forces within each PR. This gives rise to interesting dynamics that depend on intra-plug stress relaxation rates, which are not described by our equations. However, by assuming that the readjustment of forces is instantaneous, we can compute the intra-plug inter-layer forces and work out explicitly the catch-up dynamics all the way to global plug flow. 
\smallskip

\begin{figure}
\begin{center}
\includegraphics[width=4cm]{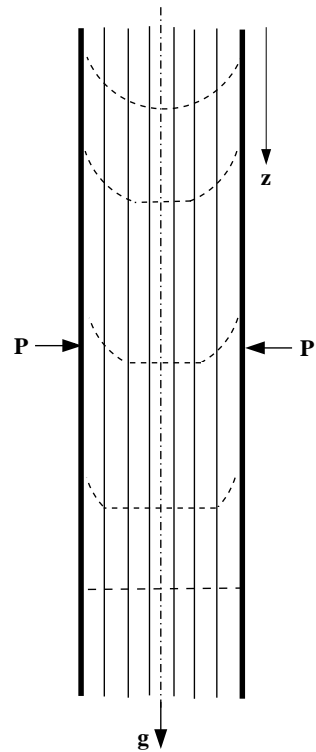}
\caption{The evolution of a velocity profile containing a local maximum. The velocity profile (dashed line) is shown at successive times from top to bottom. The fastest layer is slowed down by friction with both its neighbors, who eventually catch up with it. A plug then nucleates and expands outwards of the maximum point.}
\label{Fig3}
\end{center}
\end{figure}

\smallskip
\ni The above convergence to a uniform plug state is not necessarily the rule. 
For example, consider a flow in a pipe, tilted at an angle $\alpha$ to the horizon (figure 3). The fluid is supported by a bottom plate, whose friction with the fluid is the same as the internal friction, and is confined by a top frictionless plate. Significantly, now the inter-layer pressure varies with position, since upper layers weigh down on lower ones, leading to varying inter-layer slippage thresholds.

\begin{figure}
\begin{center}
\includegraphics[width=4cm]{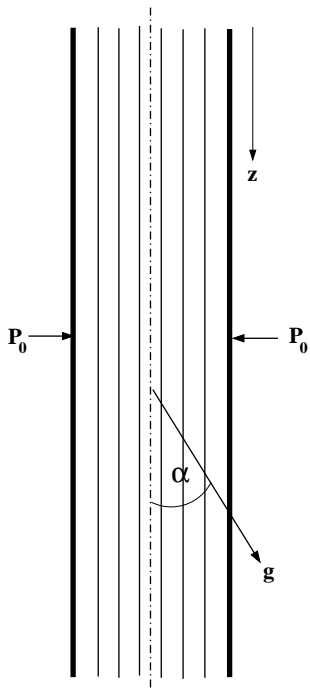}
\caption{A system of $N$ identical layers tilted to the direction of gravity $g$.}
\label{Fig4}
\end{center}
\end{figure}

\ni It is straightforward to show that, if  all layers are initially at rest under a uniform boundary pressure $P_0$, $n$ top layers slide down as one, while the bottom $N-n$ layers remain at rest. The value of $n$ depends on the angle $\alpha$ and on the friction coefficient $\mu_s$, it is the smallest integer that satisfies the relation

\begin{equation}
n\geq\frac{\mu_sP_0}{g\rho d(\cos\alpha - \mu_s\sin\alpha)}
\label{Aix}
\end{equation}
The $n$-layer PR does not expand to span the entire system because, once it starts sliding, the friction coefficient between it and the $n+1$th stationary layer drops to $\mu_d$, reducing the tangential force. This demonstrates (demonstrating) that the dVF equations accommodate stable flow solutions of localized PR's. The analysis can be extended to arbitrary initial velocity profiles. In particular, it is possible to identify the thresholds for inter-layer relative motion and the times that layers join the PR. These give rise to rich patterns of sliding regions, which will be explored elsewhere. 

\smallskip
\ni We now use the insight from discrete systems to formulate continuum flow equations. We first rewrite the discrete equations to take into account the non-differentiable form of the friction forces. 
Then we make the equations continuous by tending the layers' widths to zero, making them into streamlines. The dynamics depend non-analytically on velocity differences between layers and the equation of motion for layer $n$ can be written as

\begin{widetext}
\begin{eqnarray}
\dot{v_n} = g & + & \frac{1}{\rho d}\left[ \alpha\left( \frac{v_{n+1}-v_n}{d}\right)p_{n+1,n} - \alpha\left( \frac{v_{n}-v_{n-1}}{d} \right) p_{n,n-1} \right] \nonumber \\
& + & \frac{\mu_d}{\rho d}\left[ P_{n+1}{\rm Sign}\left( \frac{v_{n+1}-v_n}{d}\right) - P_{n-1}{\rm Sign}\left( \frac{v_{n}-v_{n-1}}{d}\right) \right] \ ,
\label{Ax}
\end{eqnarray}
\end{widetext}where we define, respectively, Sign$(u)=-1,0,1$ when $u<0,u=0,u>0$ and $\alpha(u)=0,1$ when $u\neq 0, u=0$.  The expression in the first brackets on the right hand side describes sub-threshold friction forces when layers move together. The second brackets contain the contribution from friction forces between neighbour layers at different velocities. Sub-threshold friction forces satisfy $p_{n+1,n}+p_{n-1,n}=F_{PR}$, where $F_{PR}$ is the total force on the PR boundary. 

\ni To extend eqs. (\ref{Ax}) to the continuum, we define a continuous coordinate, $x=nd$, running normal to the boundaries from left to right (figure 1) and take the limits $d\to 0$ and $N\to\infty$, such that $Nd=L$.  For generality, we let $P=P(x)$ be a function of position.  In this limit we obtain   
\begin{widetext}
\begin{equation} 
\frac{\partial v}{\partial t} = g + \frac{1}{\rho}\frac{\partial}{\partial x} \left[ p(x) \ \alpha\left(\frac{\partial v}{\partial x}\right)\right] 
+ \frac{\mu_d}{\rho}\frac{\partial}{\partial x} \left[ P(x) \ {\rm Sign}\left(\frac{\partial v}{\partial x}\right) \right]  \  , 
\label{Axi} 
\end{equation}
\end{widetext}
where $p(x)$ is the friction force per unit area  between streamlines.  
As in discrete systems, this flow is unstable to nucleation and growth of PR's, whose boundaries are where $\partial v / \partial x$ changes discontinuously from zero to a finite value. 

\ni We next obtain an equation for the boundary growth. Consider a fluid, containing at least one PR, flowing in the $z$-direction under gravity $g$ with a velocity profile $v_p(x)$ at $t=0$. Consider a PR of width $w$ between $x_l$ and $x_r=x_l+w$ moving at velocity $u_0$. The left and right streamlines just outside the PR move, say, slower than the plug and, for concreteness, we choose their initial velocities as $v_l < v_r < u_0$. The streamlines apply a deccelerating force of $2\mu_d P$ on the PR and, after a short time $t$, the velocities of the PR and its boundary streamlines are 

\begin{eqnarray}
v_r(t) & = & v_r(0)+gt  \nonumber \\      
u(t) & = & u_0+\left(g-\frac{2\mu_d P}{\rho w}\right)t  \\
v_l(t) & = & v_l(0)+gt \  . \nonumber
\label{Axii}
\end{eqnarray}
\ni The right streamline catches up with the plug first after
$\tau_r = \rho w  \left[ u_0 - v_r(0)\right] / 2\mu_d P$.
\ni Dividing both sides of this expression by $d$ and taking the continuum limit $d\to 0$, the expansion of the PR to the right follows

\begin{equation}
\frac{dx_r}{dt} = -\frac{2\mu_d P}{\rho w} \left(\frac{\partial v}{\partial x} \right)_{x_r}^{-1} \  ,
\label{Axiv}
\end{equation}
where $\left(\partial v/\partial x\right)_{x_r}$ is the gradient of the $z$-directed velocity at the right boundary. Combined with a similar expansion to the left, we find that the plug broadens at a rate

\begin{equation}
\frac{dw}{dt} = \frac{2\mu_d P}{\rho w} 
\left[ \left( \frac{\partial v}{\partial x} \right)_{x_l}^{-1} -\left( \frac{\partial v}{\partial x} \right)_{x_r}^{-1} \right] \  .
\label{Axvi}
\end{equation}
\ni Suppose the initial flow velocity profile is analytic with a local maximum at $x=0$, $v(x,t=0)=u_0 - C x^2 + \ldots$. This gives birth to a PR that broadens, according to (\ref{Axvi}), as

\begin{equation}
\frac{dw}{dt} = \frac{2\mu_d P }{\rho C w^2} \qquad \Rightarrow \qquad  w = \left(\frac{2\mu_d P}{3\rho C}t\right)^{1/3}\  .
\label{Axvii}
\end{equation}
\ni This broadening as $t^{1/3}$ is generic, occurring in different geometries and higher dimensions, as will be reported elsewhere \cite{ScBl08}. 

\bigskip
\section{Conclusion and discussion}

\ni To conclude, we have presented a minimal description of flow of dense granular matter, modeled as a da Vinci fluid. The dVF is governed by Newton's equations with normal contact forces and drag due to solid friction between volume elements. 
Consequently, accelerations cannot be ignored, as in many existing models (e.g. \cite{Je87}.
We first analysed the behavior of discrete systems, showing that the threshold nature of the da Vinci - Amontons - Coulomb friction law gives rise to rich dynamics. Most notably, we have shown that the flow is unstable to nucleation of plug regions, wherein the fluid flows at a uniform (position-independent) velocity. We have found that, under some conditions, once a PR has nucleated, it may expand. 
We have identified a growth mechanism, which we call catch-up dynamics, and illustrated the dynamics explicitly in several examples that provide clear physical insight. 

\ni We have then extended the formulation to the continuum, where the discrete layers become streamlines whose drag is governed by the da Vinci - Amontons - Coulomb friction. 
We find that, under appropriate conditions PR's expand, in which case their linear size grows proportionally to $t^{1/3}$. In these cases PR's expand and coalesce until the entire fluid flows as one plug. We have also shown that, under some boundary loading, PR's may remain finite. It is interesting that dense clusters forming in granular gases also grow as $t^{1/3}$ \cite{Third}. 

\ni The normal hysteresis associated with the different values of $\mu_s$ and $\mu_d$, hardly plays a role in the model. This is because the value of $\mu_s$ is only implicit as a threshold in the $p_{i,j}$Õs in eq. (\ref{Ax}). The hysteresis would come into play when intra-plug layers start moving relative to one another, but there is no mechanism to initiate internal motion. Forces are applied to plugs only through their boundaries, via dynamic friction. 

\ni The model has several advantages. 
First, it is minimal, involving only inertia and solid friction, and hence it is straightforward to interpret physically. In particular, the model alleviates the need to resort to an energy equation since the energy, for this type of fluid, can be computed directly from the equations. Indeed, the irrelevance of energy as a variable in linear Couette flow of granular materials has been discussed recently by Kumaran \cite{Ku08}.
Second, it gives rise generically to plug flow, a phenomenon often observed in flow of dense granular fluids, via a tractable physical mechanism. The simplicity of the model makes it useful for gaining insight into the formation and development of plug regions, and in particular into their peculiar rate of expansion.

\ni It should be noted that, in contrast to many existing models, which assume a steady state flow of constant velocity profile, there is no such assumption in the dVF model. For example, under the pure gravitational flow shown in figures 1 and 2, the steady state consists of a plug flow moving at a constant acceleration. This is a direct consequence of considering only solid friction drag. 
This also sets the range of validity of the model because, as local velocities increase under the acceleration, the mean number of contacts per grain at any moment decreases. Once this value drops below one, the material can be considered dense no longer and the flow crosses over to what is known as the inertial regime. Thus, this model, while applying strictly to dense flows, also shows how the fluid can exit the dense regime. 

\ni To keep the model minimal, we have assumed uniform friction coefficients throughout the material. This simplification gives good insight into the phenomenon of plug formation and growth, but it prevents capturing some phenomena. Notably, it does not predict shear banding and boundary pressures that vary with boundary shear rates (e.g. as in gravitational chute flows, where, given an inertial number $I$ \cite{DaCruz05}, the pressure varies as the square of the shear rate). It also leads directly to convergence of the flow to a plug that spans the entire volume. Realistically, the local effective friction coefficient may be a function of the local density, as well as of microscopic grain-scale details. For example, the local value of $\mu$ can be suppressed considerably by rolling of grains, inducing shear banding in the model. Rollability of grains depends on: local shear, grain aspect ratios and local connectivity, which is related to the local density. Therefore, a straightforward extension to model shear banding and shear-rate-dependent boundary pressure would be by letting the local friction vary with the local density. Another straightforward way to include non-uniform dynamic friction is by following phenomenological expressions suggested for $\mu_d$ \cite{Jop06}\cite{ Chevoir07}. 
These extensions will be reported elsewhere.

\ni Note that, to model dense granular fluids, whose particles are in contact all the time, we focused on quasi-static flows, where the stress relaxation in the fluid is faster than any other relevant physical process. Assuming instantaneous stress adjustment within plugs is particularly helpful for calculating the dynamics of coalescence of plugs, which has only been mentioned briefly here. It remains to be seen whether introduction of stress propagation dynamics (sound waves) is relevant to modelling dense granular flows as da Vinci Fluids.

\end{document}